# Microgrid Value of Ramping

Alireza Majzoobi, Mohsen Mahoor, Amin Khodaei
Dept. of Electrical and Computer Engineering
University of Denver
Denver, CO, USA
Alireza.Majzoobi@du.edu, Mohsen.Mahoor@du.edu, Amin.Khodaei@du.edu

*Abstract*—**The growing penetration of renewable generation in distribution networks, primarily deployed by end-use electricity customers, is changing the traditional load profile and inevitably makes supply-load balancing more challenging for grid operators. Leveraging the potential flexibility of existing microgrids, that is to help with supply-load balance locally, is a viable solution to cope with this challenge and mitigate existing net load variability and intermittency in distribution networks. This paper discusses this timely topic and determines the microgrid value of ramping based on its available reserve using a cost-benefit analysis. To this end, a microgrid ramping-oriented optimal scheduling model is developed and tested through numerical simulations to prove the effectiveness and the merits of the proposed approach in microgrid ramping valuation.**

*Index Terms*— **Microgrid, network flexibility support, optimal scheduling, value of ramping.**

## Nomenclature

*Indices:*

| | |
|---|---|
| ch | Superscript for energy storage charging mode. |
| dch | Superscript for energy storage discharging mode. |
| d | Index for loads. |
| h | Index for time periods (day). |
| i | Index for DERs. |
| t | Index for time periods (hour). |

*Sets:*

| | |
|---|---|
| D | Set of adjustable loads. |
| G | Set of dispatchable units. |
| S | Set of energy storage systems. |
| W | Set of non-dispatchable units. |

*Parameters:*

| | |
|---|---|
| DR | Ramp down rate. |
| DT | Minimum down time. |
| E | Load total required energy. |
| F(.) | Generation cost. |
| MC | Minimum charging time. |
| MD | Minimum discharging time. |
| MU | Minimum operating time. |
| R | Reserved ramping power. |
| UR | Ramp up rate. |
| UT | Minimum up time. |
| $\alpha, \beta$ | Specified start and end times of adjustable loads. |
| $\rho^M$ | Market price. |
| $\eta$ | Energy storage efficiency. |
| $\tau$ | Time period. |
| $\mu$ | Microgrid value of ramping. |

*Variables:*

| | |
|---|---|
| C | Energy storage available (stored) energy. |
| D | Load demand. |
| I | Commitment state of dispatchable units. |
| P | DER output power. |
| $P^M$ | Utility grid power exchange with the microgrid. |
| $T^{ch}$ | Number of successive charging hours. |
| $T^{dch}$ | Number of successive discharging hours. |
| $T^{on}$ | Number of successive ON hours. |
| $T^{off}$ | Number of successive OFF hours. |
| u | Energy storage discharging state (1 when discharging, 0 otherwise). |
| v | Energy storage charging state (1 when charging, 0 otherwise). |
| z | Adjustable load state (1 when operating, 0 otherwise). |

## I. Introduction

WITH the rapid growth of renewable generation technology and time-consuming process of building large-scale fast response units to provide the required flexibility, imminent bottlenecks in balancing electricity supply and demand are appearing in power systems. Present renewable portfolio standards in 27 states, net metering policy and incentive programs in 43 states, and the declining prices of renewable generation technologies are the driving forces of this radical change in the United States [1]-[2], where similar incentives and drivers can be seen around the world. The so-called "duck curve" studied by the California Independent System Operator (CAISO) is a crystallized example of this bottleneck. By increasing renewable energy generation, it is anticipated that the CAISO requires novel effective approaches to cope with large ramping which is forecasted to be in the order of more than 4 GW/hour [3].

To tackle this obstacle, system operators have traditionally reaped the benefits of large-scale power generation units, such as fast response hydro and gas-fired units that not only can be dispatched rapidly, but also are able to be ramped up quickly. However, these units have some major drawbacks, including limitations in terms of quantity and capacity, being capital- and labor-intensive, and prone to possible transmission network congestions. To address this challenge, extensive research studies are conducted with a focus on addressing the variability of renewable generation as well as materializing the forecasting efforts of renewable generation [4]. Due to the continuing proliferation of renewable generation, significant uncertainty is emerged in power system operation and planning. Thus, accurate forecasting of renewable generation, load, and market prices has become a major area of research [5]-[8]. Study in [9] proposes a real-time pricing approach considering the costs that the generation ramping imposes to the power system, consisting fuel inefficiency, aging of equipment due to fast ramping, over design of power generation units, and social costs of blackouts with higher probability. This problem is also studied from large-scale and small-scale perspectives, where large-scale methods deal with managing the generation of wind and solar farms [10], [11] while small-scale approaches take the advantages of various demand side management methods, such as demand response [12], [13], energy storage [14], [15], aggregated electric vehicles scheduling [16]. However, each of the abovementioned methods face limitations that could weaken their capabilities. Authors in [17] propose a decomposition-based approach in which the effect of ramping cost in the hourly generation scheduling of the thermal units is investigated. In [18], taking advantage of flexible resources such as thermal units, energy storage, and demand response, a stochastic day-ahead scheduling is proposed in order to manage the generation variability of renewable energy sources. The ramping capability of dispatchable generation units is formulated in [19] and its impact on day-ahead generation scheduling is investigated. Ancillary services in power systems with high penetration of renewable generation is studied in [20] and the effect of ramping value on the amount and type of power generation to mitigate the renewable generation variability is investigated.

A viable solution to locally address the required flexibility in the distribution networks is to use microgrids. Microgrids are able to provide both consumers and utility companies with considerable benefits including, but not limited to, improved resiliency, reliability, power quality, and energy efficiency. In terms of operation, microgrids can operate in both grid-connected and islanded modes. In the default operation mode, i.e., grid-connected, the microgrid has the ability to exchange power with the utility gird to supply its local loads while minimizing the operation cost. On the other hand, the microgrid has the unique capability of switching to the islanded mode in order to guarantee a reliable operation. In this case, the microgrid autonomously disconnects itself from the upstream grid whenever it detects faults or disturbances [21]-[24]. Microgrids deployments have been significantly growing in recent years where it is expected that in the near future power systems be made up of a network of interconnected microgrids [25], [26].

By leveraging potential flexibility of existing microgrids, viable schemes for addressing the challenging issue of renewable generation integration and supporting distribution grid flexibility are proposed and investigated in [27]-[29]. However, lacking is the proper valuation mechanism that determines the microgrid value of ramping and enables participation in a distribution market or utility support programs. This paper builds on the existing work on microgrid ramping but focuses on identifying the true value of the offered ramp. As proposed and modeled in this paper, the microgrid operator carries out a cost-benefit analysis to determine the value of ramping to the utility grid. This value, as will be shown in this paper, will depend on several factors, from the mix of resources that the microgrid utilizes to the number of hours that the microgrid offers ramping services to the grid.

The rest of the paper is organized as follows. Section II presents the outline of the proposed model and further formulates the problem. Section III provides the numerical simulations to show the merits of the proposed model when applied on a test system. Finally the conclusions are provided in Section IV.

## II. MODEL OUTLINE AND PROBLEM FORMULATION

In order to find the microgrid value of ramping, two problems are defined, which are explained in this section. Fig. 1 illustrates the flowchart of the proposed model. The first problem is a price-based optimal scheduling which determines the optimal schedule of all distributed energy resources (DERs) and loads as well as exchanged power with the utility grid to ensure a least-cost operation. The second problem is a ramping-oriented optimal scheduling in which an additional constraint is added to the price-based model to account for the required reserved ramping in the microgrid. In this problem, the microgrid controller manages available DERs and loads in a way that not only to supply local loads with least operation cost, but also maintains a specific amount of ramping as reserve (i.e., synchronized with the grid and available to be dispatched) for supporting the utility grid. Accordingly, the microgrid value of ramping is calculated through a comparison of the results of these two problems.

A one-year scheduling horizon is considered for the proposed optimal scheduling models. This extended scheduling horizon would provide adequate amount of data to decide on the value of ramping while at the same time consider variations in loads, generations, and prices through various days, months, and seasons.

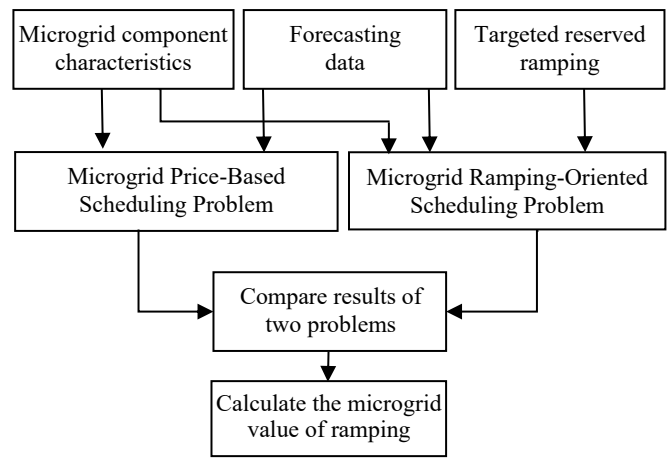

Fig. 1 Flowchart of the proposed model for calculation of microgrid value of ramping.

### A. Microgrid Component Modeling

The microgrid components to be modeled are dispatchable distributed generations (DGs), energy storage and adjustable loads. These components need to be modeled and used in both price-based and ramping-oriented microgrid optimal scheduling models, therefore proper formulation is developed below to be further used later in the paper (1)-(25):

$$\sum_i P_{iht} + P_{ht}^M = \sum_d D_{dht} \quad \forall h, \forall t, \quad (1)$$

$$-P^{M,\max} \leq P_{ht}^M \leq P^{M,\max} \quad \forall h, \forall t, \quad (2)$$

$$P_i^{\min} I_{iht} \leq P_{iht} \leq P_i^{\max} I_{iht} \quad \forall i \in G, \forall h, \forall t, \quad (3)$$

$$P_{iht} - P_{ih(t-1)} \leq UR_i \quad \forall i \in G, \forall h, t \neq 1, \quad (4)$$

$$P_{ih1} - P_{i(h-1)T} \leq UR_i \quad \forall i \in G, \forall h, \quad (5)$$

$$P_{ih(t-1)} - P_{iht} \leq DR_i \quad \forall i \in G, \forall h, t \neq 1, \quad (6)$$

$$P_{i(h-1)T} - P_{ih1} \leq DR_i \quad \forall i \in G, \forall h, \quad (7)$$

$$T_i^{on} \geq UT_i(I_{iht} - I_{ih(t-1)}) \quad \forall i \in G, \forall h, t \neq 1, \quad (8)$$

$$T_i^{on} \geq UT_i(I_{ih1} - I_{i(h-1)T}) \quad \forall i \in G, \forall h, \quad (9)$$

$$T_i^{off} \geq DT_i(I_{ih(t-1)} - I_{iht}) \quad \forall i \in G, \forall h, t \neq 1, \quad (10)$$

$$T_i^{off} \geq DT_i(I_{i(h-1)T} - I_{ih1}) \quad \forall i \in G, \forall h, \quad (11)$$

$$P_{iht} \leq P_{iht}^{dch,\max} u_{iht} - P_{iht}^{ch,\min} v_{iht} \quad \forall i \in S, \forall h, \forall t, \quad (12)$$

$$P_{iht} \geq P_{iht}^{dch,\min} u_{iht} - P_{iht}^{ch,\max} v_{iht} \quad \forall i \in S, \forall h, \forall t, \quad (13)$$

$$u_{iht} + v_{iht} \leq 1 \quad \forall i \in S, \forall h, \forall t, \quad (14)$$

$$C_{iht} = C_{ih(t-1)} - (P_{iht} u_{iht} \tau / \eta_i) - P_{iht} v_{iht} \tau$$
$$\forall i \in S, \forall h, t \neq 1, \quad (15)$$

$$C_{ih1} = C_{i(h-1)T} - (P_{ih1} u_{ih1} \tau / \eta_i) - P_{ih1} v_{ih1} \tau$$
$$\forall i \in S, \forall h, \quad (16)$$

$$C_i^{\min} \leq C_{iht} \leq C_i^{\max} \quad \forall i \in S, \forall h, \forall t, \quad (17)$$

$$T_{iht}^{ch} \geq MC_i(u_{iht} - u_{ih(t-1)}) \quad \forall i \in S, \forall h, t \neq 1, \quad (18)$$

$$T_{ih1}^{ch} \geq MC_i(u_{ih1} - u_{i(h-1)T}) \quad \forall i \in S, \forall h, \quad (19)$$

$$T_{iht}^{dch} \geq MD_i(v_{iht} - v_{ih(t-1)}) \quad \forall i \in S, \forall h, t \neq 1, \quad (20)$$

$$T_{ih1}^{dch} \geq MD_i(v_{ih1} - v_{i(h-1)T}) \quad \forall i \in S, \forall h, \quad (21)$$

$$D_d^{\min} z_{dht} \leq D_{dht} \leq D_d^{\max} z_{dht} \quad \forall d \in D, \forall h, \forall t, \quad (22)$$

$$T_d^{on} \geq MU_d(z_{dht} - z_{dh(t-1)}) \quad \forall d \in D, \forall h, t \neq 1, \quad (23)$$

$$T_d^{on} \geq MU_d(z_{dh1} - z_{d(h-1)T}) \quad \forall d \in D, \forall h, \quad (24)$$

$$\sum_{[\alpha,\beta]} D_{dht} = E_d \quad \forall d \in D. \quad (25)$$

The load balance equation (1) ensures that the sum of microgrid local DERs generation and exchanged power with the utility grid equals total local load demand. The microgrid power exchange with the utility grid is restricted to the capacity of the line between the microgrid and the utility grid (2). Hourly generation of distributed DG units is limited to the maximum and minimum capacity of the units (3), where the unit commitment state variable $I$ defines the commitment state of each unit, i.e. $I=1$ when the unit is ON and zero otherwise. Ramp up and ramp down constraints of dispatchable DG units are defined by (4)-(7), where (4),(6) and (5),(7) belong to intra-day and inter-day intervals, respectively. Constraints (8)-(11) represent the minimum up and down time limits of dispatchable DG units, where (8) and (10) belong to intra-day intervals, while (9) and (11) belong to inter-day intervals. The limitation of energy storage power depends on the charging and discharging minimum and maximum limits, based on its mode (12)-(13). In the charging mode the binary charging variable $v$ is one and the binary discharging variable $u$ is zero, and vice versa. In addition, (14) ensures that the energy storage can only operate in one mode, i.e., either charging or discharging, at every given time period. Constraints (15) and (16) respectively define the amount of available stored energy in the energy storage, in intra-day and inter-day intervals, which is further confined by the energy storage capacity (17). Constraints (18) and (19) represent the minimum charging time of energy storage for intra-day and inter-day intervals, respectively. Similarly, the maximum charging time of energy storage for intra-day and inter-day intervals are respectively defined by (20) and (21). Constraint (22) represents the minimum and maximum rated power of adjustable loads, where the binary operating state of adjustable load, i.e., $z$, is one when load is consuming power, and it is zero otherwise. The minimum operating time of adjustable loads for intra-day and inter-day intervals are defined by (23) and (24), respectively. Moreover, each load consumes the required energy to complete an operating cycle in time intervals specified by consumers, where $\alpha$ and $\beta$ respectively represent the start and end operating time of loads (25). It is worthwhile to mention that $T$ represents the last hour of a day, i.e. $T=24$.

These components are modeled for each day of operation in one-year scheduling horizon, and furthermore, for each hour of operation within each day. This modeling ensures that daily schedules and habits can be taken into account, such as operation of adjustable loads and charge/discharge schedule of energy storage. Relevant constraints are defined for inter-day intervals, to connect the operation schedule at last hour of operation in each day to the first hour of operation in its next.

*B. Price-based and Ramping-oriented Optimal Scheduling*

In the price-based optimal scheduling, the microgrid is seeking a minimum operation cost as formulated in the following:

$$\min \sum_h \sum_t [\sum_{i \in G} F_i(P_{iht}) + \rho_{ht}^M P_{ht}^M], \quad (26)$$

subject to (1)-(25).

The first term in the objective function (26) is operation cost of dispatchable units and the second term is the revenue or expense of microgrid through power exchange with the utility grid. When the microgrid is purchasing energy from the utility grid, $P^M$ is positive, and when the microgrid is selling its excess energy to the utility grid, $P^M$ is negative, respectively representing a cost and a benefit for the microgrid.

In the ramping-oriented microgrid optimal scheduling, the microgrid not only is responsible for supplying its local loads, but also provides the required ramping to the utility grid. The proposed model is formulated as following:

$$\min \sum_h \sum_t [\sum_{i \in G} F_i(P_{iht}) + \rho_{ht}^M P_{ht}^M] - \mu \sum_h \sum_t R_{ht}, \quad (27)$$

subject to (1)-(25), and

$$\sum_{i \in G} P_i^{\max} I_{iht} + \sum_{i \in \{S,W\}} P_i + P_{ht}^M \geq \sum_d D_{dht} + R_{ht} \quad \forall h, \forall t. \quad (28)$$

The objective function is similar to what is used in the price-based optimal scheduling model, however it has an additional term that represents the ramping cost. In the ramping cost, $R$ is the amount of ramping that the microgrid can offer (i.e., reserved power) and $\mu$ is the microgrid value of ramping ($/MWh).

In order to consider the reserved power in the microgrid to support distribution network flexibility, (28) is developed and added to this problem. In this constraint, the summation of the maximum capacity of dispatchable DG of committed units, power generation of non-dispatchable units, and exchanged power with the utility grid should be greater than sum of microgrid loads (fixed and adjustable) and reserved power at each hour. So, (28) ensures that the total power generation of local microgrid DERs and exchanged power with the utility grid not only supplies microgrid local demand, but also at least extra power ($R$) is reserved at the desired time intervals in order to support distribution network flexibility. The reserved power ($R$) is considered a time-dependent parameter in the model which gives the ability of considering the reserved ramping capability in any desired time interval, or a series of time intervals, within the scheduling horizon.

### C. Microgrid Value of Ramping Calculation

To find the microgrid value of ramping, the two abovementioned problems are solved and the solutions are compared to find value of ramping. The difference between microgrid operation cost in these two problems is the extra cost which is imposed to the microgrid, owing to considering $R$ (MW) reserved ramping to support the utility grid. Therefore, at least the amount of $\mu R$, aggregated over all time intervals in the scheduling horizon, should be paid to the microgrid for maintaining unused capacity to offer requested ramping by the utility grid. In other words, the minimum value of ramping is determined as in (29):

$$\mu^{\min} = (C^{RO} - C^{PB}) / \sum_h \sum_t R_{ht}. \quad (29)$$

where $C^{RO}$ represents the objective value of ramping-oriented microgrid optimal scheduling problem and $C^{PB}$ represents the objective value of price-based microgrid optimal scheduling problem.

### III. NUMERICAL EXAMPLES

A microgrid with four dispatchable units, two nondispatchable units (wind and solar), one energy storage, and five adjustable loads is utilized for studying the performance of the proposed model. The details of microgrid DERs are borrowed from [30], and annual data for hourly market price, load, wind and solar generation are borrowed from [31]. A maximum ramping capability of 18 MW/h is available in the microgrid, based on the maximum ramping capacity of dispatchable DGs. In addition, a limit of 10 MW is considered as the maximum capacity of the line connecting the microgrid to the utility grid. The developed mixed-integer programming problems are solved using CPLEX 12.6, with a computation time between 5 and 6 minutes for each studied case. The following five cases are investigated:

**Case 1:** Price-based optimal scheduling
**Case 2:** Price-based optimal scheduling, considering a 2 MW reserved ramping capability in all operation hours.
**Case 3:** Price-based optimal scheduling, considering a 2 MW reserved ramping capability in all operation hours, along with uncertainty on load and nondispatchable generation.
**Case 4:** Price-based optimal scheduling, considering a 2 MW reserved ramping capability in specific operation hours, along with uncertainty on load and nondispatchable generation.
**Case 5:** Sensitivity analysis of value of ramping with respect to the amount of reserved ramping capability.

**Case 1:** The grid-connected price-based optimal scheduling problem is solved for the considered one-year horizon as a base case. In this case, the microgrid is only responsible for minimizing its operation cost via managing its dispatchable generation units and adjustable load, and does not have any commitment to the utility grid in terms of ramping. The microgrid operation cost in this case is calculated as $1,720,193.

**Case 2:** In this case, the microgrid not only is responsible for minimizing its operation cost, but also commits 2 MW as the ramping for supporting the utility grid in all operation hours in the scheduling horizon. Fig. 2 depicts the microgrid exchanged power with the utility grid in Cases 1 and 2 in a sample day of the studied year. To realize the microgrid behavior in power arbitrage in various hours of a day, the market price is also shown in this figure.

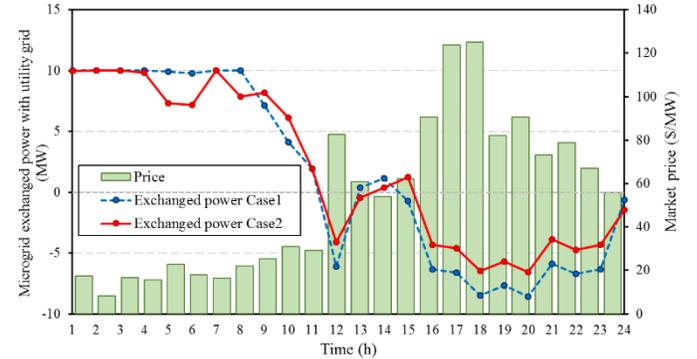

Fig. 2. Microgrid exchanged power with the utility grid in a sample day of the studied year, in Case 1 and Case 2.

As this figure shows, in Case 1 the microgrid buys power from the utility grid in full capacity from midnight to early morning, when the market price is the lowest. Then, in the morning, with increasing the market price, microgrid reduces its import power from the utility grid and even at noon it sells excess power back to the utility grid. Again, from early evening to midnight (hours 15 to 24), when the market price is high, microgrid sells its excess power to the utility grid in order to increase its revenue. Thus, in price-based optimal scheduling, microgrid maximizes its revenue via managing its local resources and power exchange with the utility grid. The general trend of microgrid power arbitrage with the utility grid in Case 2 is almost the same as Case 1, since in this case still microgrid aims at minimizing its operation cost. But in addition to minimizing its operation cost, 2 MW is considered as reserved power which means that microgrid has the capability of offering up to 2 MW/h ramping to the utility grid in all hours during the scheduling horizon. The microgrid operation cost in Case 2 increases to $1,901,963 (10.5% increase compared with Case 1), in expense of offering the ramping service to the utility grid. Furthermore, the microgrid optimal schedule is changed as microgrid sells less power in the afternoon and evening hours, which means a smaller revenue for the microgrid. The difference of microgrid operation cost in these two cases can be used to find the value of ramping, as in (29), which in this case is calculated as $10.4/MWh. This is the minimum price that should be offered to the microgrid in order to maintain 2 MW reserve ramping in all operation hours within a one-year scheduling horizon.

**Case 3:** In this case, in addition to considering 2 MW reserved power in all operation hours, +10% and -20% forecast error in load and nondispatchable generation are respectively considered. Since the forecast error in load, solar, and wind generation is inevitable, considering these uncertainties make a more practical case. Furthermore, considering +10% uncertainty for load and -20% for solar/wind generation is the worst-case scenario for the model to be sure that the microgrid will have the capability of offering 2 MW ramping, even if its solar and wind generation drop by 20% and/or the load increases by 10%. The microgrid operation cost in this case increases to $2,397,799, which is 39% more than Case 1 due to considering 2 MW reserved power along with uncertainty, and 26% more than Case 2 owing to adding aforementioned uncertainties. As the microgrid operation cost in the base case (price-based optimal scheduling) while considering uncertainty is $2,224,390, the value of ramping for the microgrid in this case is equal to $9.89/MWh.

Fig. 3 compares microgrid exchanged power with the utility grid in Cases 2 and 3. As the figure illustrates, considering uncertainty leads to changes in microgrid optimal schedulling. Since the generation of nondispatchable units have been decreased and the local load of microgrid has been increased, in order to still keep 2 MW reserved power, the microgrid buys more power and sells less in all operation hours.

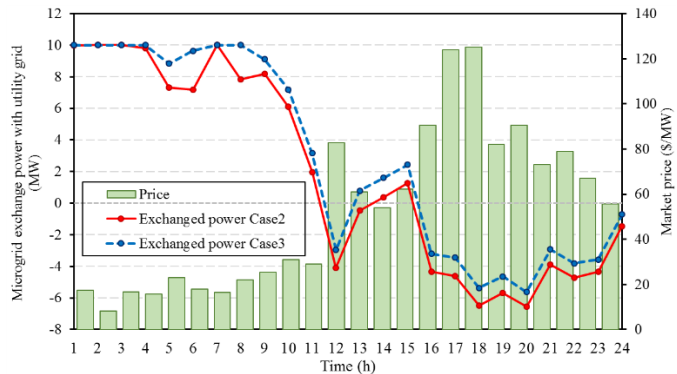

Fig. 3. Microgrid exchanged power with the utility grid in a sample day of the studied year, in Case 2 and Case 3.

**Case 4:** In this case, instead of considering reserved power in all operation hours of a year, it is only considered for the specific hours, specifically at times that the utility grid demand for ramping might be higher. To this end, three hours in 30 different days of a year (mainly peak load hours) are selected for considering a 2 MW reserved power. The microgrid operation cost in this case is $2,230,500 which is $6,110 more than microgrid operation cost without consideration of reserved power. Hence, the value of ramping is calculated as $33.9/MWh.

**Case 5:** In this case, the sensitivity of microgrid value of ramping with respect to the amount of reserved ramping is analyzed. Fig. 4 depicts the microgrid value of ramping for various amounts of reserved power (MW) in two different scenarios: (i) considering reserved ramping for all hours of the year, (ii) considering reserved ramping for only 90 hours of the year.

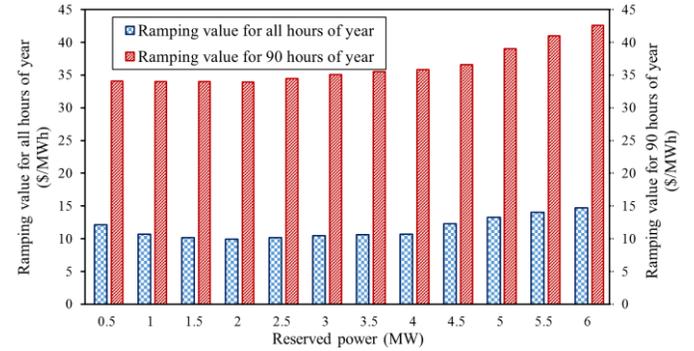

Fig. 4. Microgrid ramping value for various amount of reserved ramping capacity and considering uncertainty.

As the figure shows, by increasing the amount of reserved power, the value of ramping slightly decreases (for less than 1 MW for each step) and after that it increases in both scenarios. In addition, the results show the microgrid value of ramping for the lower number of the hours in a year is higher. Because offering ramping services to the utility grid for lower number of hours leads to less total revenue for the microgrid, so the higher value of ramping will compensate the smaller number of hours to make reasonable total revenue for microgrid in order to participate in a distribution market or utility support programs.

TABLE I
MICROGRID OPERATION COST ($)

| Reserved ramping capacity (MW) | Without Uncertainty | | With Uncertainty | |
|---|---|---|---|---|
| | Reserved ramping for all year | Reserved ramping for 90 hours | Reserved ramping for all year | Reserved ramping for 90 hours |
| 0.0 | 1,720,193 | 1,720,193 | 2,224,390 | 2,224,390 |
| 0.5 | 1,782,740 | 1,721,723 | 2,277,466 | 2,225,923 |
| 1.0 | 1,822,735 | 1,723,249 | 2,317,815 | 2,227,449 |
| 1.5 | 1,862,287 | 1,724,775 | 2,357,755 | 2,228,975 |
| 2.0 | 1,901,963 | 1,726,301 | 2,397,799 | 2,230,500 |
| 2.5 | 1,950,841 | 1,727,950 | 2,446,816 | 2,232,150 |
| 3.0 | 2,002,903 | 1,729,666 | 2,498,886 | 2,233,866 |
| 3.5 | 2,052,490 | 1,731,383 | 2,549,218 | 2,235,583 |
| 4.0 | 2,101,422 | 1,733,099 | 2,599,055 | 2,237,299 |
| 4.5 | 2,218,898 | 1,735,017 | 2,709,186 | 2,239,217 |
| 5.0 | 2,313,720 | 1,737,742 | 2,805,030 | 2,241,942 |

Moreover, the microgrid operation cost in various cases, including reserved ramping for all hours and 90 hours of the year, with and without uncertainty, are tabulated in Table I. The obtained results in Table I demonstrate that the microgrid operation cost increases by augmenting reserved ramping capacity, in all cases. In addition, the results prove the significant effect of considering uncertainty on microgrid operation cost. In both conditions of considering reserved power for all hours and 90 hours of the year, uncertainty imposes between 20% and 30% increase on the microgrid operation cost, depending on the amount of considered reserved power. It is worth mentioning that the zero value for reserved power in the table represents the price-based optimal scheduling of the microgrid, i.e. base case.

IV. CONCLUSIONS

In this paper, a microgrid ramping-oriented scheduling model was proposed to offer the flexibility support to the utility grid. The microgrid value of ramping was further calculated through comparing the results of microgrid optimal scheduling and microgrid ramping-oriented scheduling problems. The obtained value of ramping could be a decisive factor for microgrid operator to whether participate in supporting distribution network flexibility or not. Numerical simulations were performed for various situations, considering reserved ramping for all hours and just specific hours of a year as well as different amounts of reserved ramping, to advocate the merits and effectiveness of the proposed model. In addition, the uncertainty on the load and renewable generation were considered in the simulations, as a worst-case. The results demonstrate that the microgrid can calculate its value of ramping, in different situations via the proposed model, in order to have an accurate and reasonable bid for participating in the distribution market or directly supporting the utility grid.